\documentclass{ifacconf}

\usepackage{graphicx}      
\usepackage{natbib}        

\usepackage{amsmath}
\usepackage{amssymb}
\usepackage{color}
\usepackage{transparent}
\usepackage{pgfplots}
\usepackage{tikz}
\usetikzlibrary{decorations.pathreplacing}
\graphicspath{{./fig/}}

\begin{document}
\begin{frontmatter}

\title{Structure preserving discontinuous Galerkin approximation of one-dimensional port-Hamiltonian systems}

\author[first]{Tobias Thoma},
\author[first]{Paul Kotyczka}

\address[first]{Technical University of Munich, TUM School of Engineering and Design, Chair of Automatic Control, Garching, Germany (e-mail: tobias.thoma@tum.de, kotyczka@tum.de).}

\thanks[footnoteinfo]{This paper has been submitted to the 22nd World Congress of the International Federation of Automatic Control.}

\begin{abstract}                
	In this article, we present the structure-preserving discretization of linear one-dimensional port-Hamiltonian (PH) systems of two conservation laws using discontinuous Galerkin (DG) methods. We recall the DG discretization procedure which is based on a subdivision of the computational domain, an elementwise weak formulation with up to two integration by parts, and the interconnection of the elements using several numerical fluxes. We present the interconnection of the element models, which is power preserving in the case of conservative (unstabilized) numerical fluxes, and we set up the resulting global PH state space model. We discuss the properties of the obtained models, including the effect of the flux stabilization parameter on the spectrum. Finally, we show simulations with different parameters for a boundary controlled linear hyperbolic system.
\end{abstract}


\end{frontmatter}

\section{Introduction}
\label{sec:1}
Port-Hamiltonian (PH) systems provide a framework for modeling, analysis and control of complex dynamical systems \citep{Duindam2009} where the complexity might result from multi-physical couplings, non-trivial domains and nonlinearity. Since modeling many engineering problems requires handling partial differential equations (PDEs), the PH community has been heavily involved in the theory of infinite-dimensional PH \citep{jacob2012linear} systems in recent years \citep{Rashad2020}. In order to solve these PDEs or generate finite-dimensional PH systems, several structure preserving discretization methods have been developed, e.g. \cite{Trenchant2018,kotyczka2019numerical,Cardoso-Ribeiro2020}. These discretization techniques preserve the characteristics of infinite-dimensional PH systems and transform them to finite-dimensional ones.

'Classical' spatial discretization methods can be classified in the categories of the finite element method (FEM), finite volume method (FVM) or finite difference method (FDM). For more information on these general methods the reader is referred to \cite{Zienkiewicz2005,Kitamura2020,Chowdhury2018}. Each of these methods is justified and has advantages and disadvantages. For example, the FDM -- one of the simplest and oldest methods -- is very intuitive and computationally efficient but limited to simple grids. In contrast, the FVM has a significantly higher geometric flexibility, but is difficult to implement as a higher-order method. The FEM -- a common tool in structural mechanics -- is geometrically flexible and can easily be extended to a higher order method. Despite its many advantages, FEM is often unsuitable for the simulation of transport phenomena or conservation laws, where the direction of information flow is well known, and leads to stability problems. In order to overcome this and to combine the advantages of FVM and FEM, the so called discontinuous Galerkin (DG) \citep{Hesthaven2008} methods were developed.

Like the FEM, DG methods are based on a weak formulation and a subdivision of the computational domain into several finite elements. Unlike the FEM, the weak form of the DG methods is not set up globally but for each individual element. Inserting approximations of trial and test functions in the weak form leads to ordinary differential equations of each element. The similarity to the FVM, is demonstrated by the interconnection of the elements what is not done by an assembly process, like in the FEM, but by numerical fluxes defined at the boundary of each element. The choice of the numerical flux has a decisive influence on the simulation accuracy. Due to this process, DG methods generate block diagonal mass matrices, being very efficient, when it comes to explicit time integration. A serious drawback of DG methods compared to FEM is the increased number of unknowns being a result of the discontinuity between elements. However, DG methods usually lead to very sparse matrices, which facilitates parallelization, whereby the disadvantage of large system orders is somewhat weakened. Moreover, the numerical flux formulation allows easy handling of hanging nodes.

One concept for discretizing linear Hamiltonian hyperbolic systems using a DG method is given in \cite{Xu2008}, where an energy conservative discretization due to special numerical fluxes is represented. Nevertheless, the article does not concern PH systems and their control oriented state space representation with boundary in- and power-conjugated outputs.

In this article, we demonstrate how DG methods can be applied to one-dimensional PH systems of conservation laws in a structure preserving way. To this end, the computational domain is divided into several elements and the weak form of each element is generated. After applying integration by parts to the first equation once and to second one twice, the numerical flux definition interconnects elements and generates a global explicit PH state space system with mixed boundary control inputs.

This article is organized as follows. In Section \ref{sec:2}, we recall the most important characteristics of one-dimensional wave-like hyperbolic systems written in a PH formulation. Section \ref{sec:3} shows the discretization procedure which yields finite-dimensional PH models with possible dissipation due to the choice of the numerical flux. In Section \ref{sec:4}, we discuss the properties of the resulting models on a numerical example, and Section \ref{sec:5} gives a short conclusion.

\section{One-dimensional PH systems of conservation laws}
\label{sec:2}
We consider linear one-dimensional parameter distributed PH systems defined on the constant spatial domain ${\Omega=[a,b]\subset\mathbb{R}}$. The system representation is based on three components.
The structure equations
	\begin{equation}
		\label{eq:pde}
		\begin{bmatrix}
			f_1(z,t) \\ f_2(z,t)
		\end{bmatrix} = 
		\begin{bmatrix}
			0 & \partial_z \\ \partial_z & 0
		\end{bmatrix}
		\begin{bmatrix}
			e_1(z,t) \\ e_2(z,t)
		\end{bmatrix}
	\end{equation}
represent the relations between the flow and effort functions ${f_1,f_2:\Omega\times\mathbb{R}\rightarrow\mathbb{R}}$ and ${e_1,e_2:\Omega\times\mathbb{R}\rightarrow\mathbb{R}}$. The second component (dynamics)
	\begin{equation}
		\label{eq:dynamics}
		\begin{bmatrix}
			\partial_t x_1(z,t) \\ \partial_t x_2(z,t)
		\end{bmatrix} = 
		\begin{bmatrix}
			-f_1(z,t) \\ -f_2(z,t)
		\end{bmatrix}
	\end{equation}
describes the evolution of the states ${x_1,x_2:\Omega\times\mathbb{R}\rightarrow\mathbb{R}}$. In order to close the system representation, the constitutive equations
	\begin{equation}
		\label{eq:constitutive}
		\begin{bmatrix}
			e_1(z,t) \\ e_2(z,t)
		\end{bmatrix} = 
		\begin{bmatrix}
			\delta_{x_1} H(x_1(z,t),x_2(z,t)) \\ \delta_{x_2} H(x_1(z,t),x_2(z,t))
		\end{bmatrix}
	\end{equation}
are required, where ${\delta_{x_j}H}$ denotes the variational derivative of the total energy/Hamiltonian functional $H$ with respect to $x_j$.
Partial derivatives are denoted by ${\partial_ze_j:=\frac{\partial e_j}{\partial z}}$.
For compact notation, we usually omit the arguments ${t\in\mathbb{R}}$ and ${z\in\Omega}$ of the occurring functions.

The efforts or co-energy variables are derived from $H$ which is assumed quadratic in this article,
	\begin{equation}
		H = \int_{\Omega}\mathcal{H}\;\text{d}z = \int_{\Omega}\frac{1}{2}c_1x_1^2 + \frac{1}{2}c_2x_2^2 \;\text{d}z,
	\end{equation}
with the constant coefficients ${c_1\in\mathbb{R}^+}$ and ${c_2\in\mathbb{R}^+}$. Since $\mathcal{H}$ does not depend on spatial derivatives of $x_1$, $x_2$, the identity
	 \begin{equation}
	 	\begin{bmatrix}
	 		\delta_{x_1} H \\ \delta_{x_2} H
	 	\end{bmatrix} = 
	 	\begin{bmatrix}
	 		\partial_{x_1} \mathcal{H} \\ \partial_{x_2} \mathcal{H}
	 	\end{bmatrix}
	 \end{equation}
holds. Taking the time derivative of the total energy, using \eqref{eq:dynamics} and \eqref{eq:constitutive}, leads to the power balance
	\begin{equation}
		\label{eq:power_balance}
		\begin{split}
		\frac{\text{d}H}{\text{d}t} &= \int_{\Omega}\partial_{x_1}\mathcal{H}\:\partial_tx_2 + \partial_{x_1}\mathcal{H}\:\partial_tx_2\;\text{d}z \\
		&= -\int_{\Omega} e_1f_1 + e_2f_2\;\text{d}z,
		\end{split}
	\end{equation}
which demonstrates the meaning of flows and efforts as dual power variables.
Inserting \eqref{eq:pde} in \eqref{eq:power_balance} and applying integration by parts to one of the addends reveals the boundary energy flow
	\begin{equation}
		\label{eq:power}
		\frac{\text{d}H}{\text{d}t} = \underbrace{-\left[e_1e_2\right]^b_a}_{-\int_{\partial\Omega}ne_2e_1\;\text{d}s}.
	\end{equation}

\begin{rem}
	The right hand side of \eqref{eq:power} results from the integration by parts in one dimension. The general form valid for all dimensions is the boundary integral over ${\partial\Omega}$ containing the outer unit normal vector $n$. In the case of one-dimensional PDEs the normal vector corresponds to ${n=1}$ on the right boundary of $\Omega$ and ${n=-1}$ on the left one.
\end{rem}

The boundary ${\partial\Omega=\{a,b\}=\Gamma_D\cup\Gamma_N}$ can be split into two subsets, where different boundary conditions are imposed:
	\begin{equation}
		\label{eq:bc}
		e_1 = u_1 \quad \text{on} \; \Gamma_D \qquad \text{or} \qquad ne_2 = u_2 \quad \text{on} \; \Gamma_N.
	\end{equation}
By considering the boundary conditions as inputs, we obtain the power-conjugated outputs ${y_1=-ne_2}$ on $\Gamma_D$ and ${y_2=-e_1}$ on $\Gamma_N$ according to
	\begin{equation}
		\frac{\text{d}H}{\text{d}t} = [u_1y_1]_{\Gamma_D} + [u_2y_2]_{\Gamma_N}.
	\end{equation}

\section{Structure preserving discontinuous Galerkin Approximation}
\label{sec:3}
We demonstrate the discretization procedure based on a DG method starting with the weak DG form.

\subsection{Weak and "strong" DG form}
\begin{figure}[htbp]
	\centering
	\begin{tikzpicture}
		\draw[-to](0,0) -- (6,0);
		\node () at (6.3,0) {$z$};
	
		\draw (1,-0.2) -- (1,0.2);
		\draw (3,-0.2) -- (3,0.2);
		\draw (5,-0.2) -- (5,0.2);
		
		\node () at (1,-0.5) {$z_{i-1}$};
		\node () at (3,-0.5) {$z_{i}$};
		\node () at (5,-0.5) {$z_{i+1}$};
		
		\draw [decorate,decoration={brace,amplitude=10pt,raise=4pt},yshift=0pt](3,-0.1) -- (5,-0.1) node [black,midway,yshift=0.7cm] {
			$\Omega_i$};
	\end{tikzpicture}
	\caption{Subdivision of the computational domain}
	\label{fig:domain}
\end{figure}
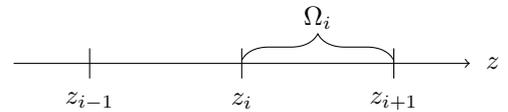

Similarly to the FEM, the DG method requires a subdivision of the computational domain $\Omega$ into a set of ${N\in\mathbb{N}^+}$ elements $\Omega_i$ with ${\Omega=\bigcup_i\Omega_i}$, see Figure \ref{fig:domain}.

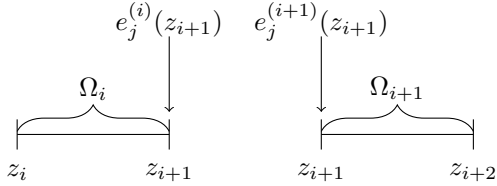
\begin{figure}[htbp]
	\centering
	\begin{tikzpicture}
		
		\draw (0,0) -- (2,0);
		\draw (0,-0.2) -- (0,0.2);
		\draw (2,-0.2) -- (2,0.2);
		
		\draw (4,0) -- (6,0);
		\draw (4,-0.2) -- (4,0.2);
		\draw (6,-0.2) -- (6,0.2);
		
		\node () at (0,-0.5) {$z_{i}$};
		\node () at (2,-0.5) {$z_{i+1}$};
		\node () at (4,-0.5) {$z_{i+1}$};
		\node () at (6,-0.5) {$z_{i+2}$};
		
		\draw [decorate,decoration={brace,amplitude=10pt,raise=4pt},yshift=0pt](0,-0.1) -- (2,-0.1) node [black,midway,yshift=0.7cm] {
			$\Omega_i$};
		\draw [decorate,decoration={brace,amplitude=10pt,raise=4pt},yshift=0pt](4,-0.1) -- (6,-0.1) node [black,midway,yshift=0.7cm] {
			$\Omega_{i+1}$};
		
		\node () at (2,1.5) {$e_j^{(i)}(z_{i+1})$};
		\draw[->] (2,1.3) -- (2,0.3);
		
		\node () at (4,1.5) {$e_j^{(i+1)}(z_{i+1})$};
		\draw[->] (4,1.3) -- (4,0.3);
	\end{tikzpicture}
	\caption{Neighboring elements and notation used in the numercal fluxes, $j\in\{1,2\}$.}
	\label{fig:elements}
\end{figure}

In contrast to the FEM, where the weak form is defined over the entire domain, it is established for each individual element in the DG method. Accordingly, the weak form of each element is given by
	\begin{subequations}
		\begin{align}
			-\int_{\Omega_i}w_1\partial_tx_1\;\text{d}z &= \int_{\Omega_i}w_1\partial_ze_2\;\text{d}z \\
			-\int_{\Omega_i}w_2\partial_tx_2\;\text{d}z &= \int_{\Omega_i}w_2\partial_ze_1\;\text{d}z,
		\end{align}
	\end{subequations}
where ${w_1,w_2:\Omega\rightarrow\mathbb{R}}$ represent the test functions. Using integration by parts leads to 
	\begin{subequations}
		\label{eq:weak_elem}
	\begin{align}
		-\int_{\Omega_i}w_1\partial_tx_1\;\text{d}z &= -\int_{\Omega_i}\partial_zw_1e_2\;\text{d}z + \left[w_1e_2\right]_{z_i}^{z_{i+1}} \\
		-\int_{\Omega_i}w_2\partial_tx_2\;\text{d}z &= -\int_{\Omega_i}\partial_zw_2e_1\;\text{d}z + \left[w_2e_1\right]_{z_i}^{z_{i+1}}.
	\end{align}
	\end{subequations}
This equation is called weak DG form.
With ${(\cdot,\cdot)_{\Omega_i}}$ denoting the bilinear form over $\Omega_i$ with one slot for the test function and one for the solution, the weak DG form can be written as
	\begin{subequations}
	\begin{align}
		-\left(w_1, \partial_tx_1\right)_{\Omega_i} &= -\left(\partial_zw_1, e_2\right)_{\Omega_i} + \left[w_1e_2\right]_{z_i}^{z_{i+1}} \\
		-\left(w_2, \partial_tx_2\right)_{\Omega_i} &= -\left(\partial_zw_2, e_1\right)_{\Omega_i} + \left[w_2e_1\right]_{z_i}^{z_{i+1}}.
	\end{align}
	\end{subequations}
Up to this point, the elements are not connected and no information is taken from neighboring elements. Since this information is required to get a global solution, the boundary terms are replaced by a general expression, the so called numerical flux, indicated by a star,
	\begin{subequations}
		\label{eq:weak}
	\begin{align}
		\label{eq:weak_a}
		\left(w_1, \partial_t x_1\right)_{\Omega_i} &= \left(\partial_zw_1, e_2\right)_{\Omega_i} - \left[w_1e_2^*\right]_{z_i}^{z_{i+1}} \\
		\label{eq:weak_b}
		\left(w_2, \partial_t x_2\right)_{\Omega_i} &= \left(\partial_zw_2, e_1\right)_{\Omega_i} - \left[w_2e_1^*\right]_{z_i}^{z_{i+1}}.
	\end{align}
	\end{subequations}
Similarly to the FVM, numerical fluxes contain information from neighboring elements, indicated by an upper index (see also Figure \ref{fig:elements}),
	\begin{align}
		e_1^*(z_i) &= e_1^*\left(e_1^{(i-1)}(z_i),e_1^{(i)}(z_i), e_2^{(i-1)}(z_i),e_2^{(i)}(z_i)\right) \\
		e_2^*(z_i) &= e_2^*\left(e_1^{(i-1)}(z_i),e_1^{(i)}(z_i), e_2^{(i-1)}(z_i),e_2^{(i)}(z_i)\right)
	\end{align}
and its choice has a massive influence on the discrete system, which is part of the next subsection.

In the next step, another integration by parts is performed on either \eqref{eq:weak_a} or \eqref{eq:weak_b}. This allows us to generate the skew-symmetric energy matrix of the finite-dimensional system. In the following, integration by parts is applied to \eqref{eq:weak_b}, what leads to
	\begin{subequations}
		\label{eq:strong}
	\begin{align}
		\label{eq:strong_a}
		\left(w_1, \partial_t x_1\right)_{\Omega_i} &= \left(\partial_zw_1, e_2\right)_{\Omega_i} - \left[w_1e_2^*\right]_{z_i}^{z_{i+1}} \\
		\label{eq:strong_b}
		\left(w_2, \partial_t x_2\right)_{\Omega_i} &= -\left(w_2, \partial_ze_1\right)_{\Omega_i} - \left[w_2(e_1^*-e_1)\right]_{z_i}^{z_{i+1}},
	\end{align}
	\end{subequations}
where \eqref{eq:strong_b} is called "strong" DG form.

\begin{rem}
	The "strong" DG form does not strongly impose continuity between elements but weakly penalizes jumps in the solution via the difference ${e_1^*-e_1}$ \citep{Hesthaven2008}.
\end{rem}

\subsection{Numerical fluxes}
As already mentioned, the individual elements must be interconnected with the aid of the numerical flux. In the following, we consider the numerical flux
	\begin{align}
		\label{eq:flux_a}
		e_1^{(i)*}(z_{i}) &= (1-\beta)e_1^{(i-1)} + \beta e_1^{(i)} + \tau(e_2^{(i-1)} - e_2^{(i)}) \\
		e_2^{(i)*}(z_{i}) &= \beta e_2^{(i-1)} +   (1-\beta)e_2^{(i)} + \xi(e_1^{(i-1)} - e_1^{(i)}) \\
		e_1^{(i)*}(z_{i+1}) &= (1-\beta)e_1^{(i)} + \beta e_1^{(i+1)} + \tau(e_2^{(i)} - e_2^{(i+1)}) \\
		\label{eq:flux_d}
		e_2^{(i)*}(z_{i+1}) &= \beta e_2^{(i)} +   (1-\beta)e_2^{(i+1)} + \xi(e_1^{(i)} - e_2^{(i+1)})
	\end{align}
with the constants ${\beta\in[0,1]}$, $\tau\in\mathbb{R}^{+}_0$ and $\xi\in\mathbb{R}^{+}_0$
which generates a discrete PH formulation of each element and the global system. These numerical fluxes represent a generalization of several common flux formulations listed in Table \ref{tab:flux}.

\begin{table}[htbp]
	\centering
	\caption{Common numerical fluxes}
	\label{tab:flux}
	\begin{tabular}{ l l }
		\hline
		Numerical flux & Parameters  \\
		\hline
		Central flux & $\beta=0.5$, $\tau=\xi=0$ \\
		Upwind flux & $\beta=\{0,1\}$, $\tau=\xi=0$ \\
		Lax-Friedrichs flux &  $\beta=0.5$, $\tau=\xi>0$ \\
		\hline
	\end{tabular}
\end{table}

In case of $\tau>0$ or $\xi>0$, numerical damping is introduced in the system, whereas ${\tau=\xi=0}$ leads to a conservative behavior \citep{Xu2008}. The influence of $\beta$, $\xi$, $\tau$ on the discrete PH state space model becomes clearer in Section \ref{sec:3.3}.

\subsection{Element discrete PH formulation}
\label{sec:3.3}
\begin{thm}
	\label{thm:fem}
	By inserting \eqref{eq:flux_a}-\eqref{eq:flux_d} into \eqref{eq:strong}, applying a Galerkin discretization on \eqref{eq:strong} and using trial and test functions from the same bases\footnote{Variables with a hat represent discrete degrees of freedom.}
	\begin{align*}
		x_1^{(i)}(z,t) \approx \phi^T(z)\hat{x}_1(t), \qquad & x_2^{(i)}(z) \approx \psi^T(z)\hat{x}_2(t), \\
		e_1^{(i)}(z,t) \approx \phi^T(z)\hat{e}_1(t), \qquad & e_2^{(i)}(z) \approx \psi^T(z)\hat{e}_2(t), \\
		w_1^{(i)}(z,t) \approx \phi^T(z)\hat{w}_1, \qquad & w_2^{(i)}(z) \approx \psi^T(z)\hat{w}_2,
	\end{align*}
	the linear PH state space model of one element
	\begin{align}
		\label{eq:element}
		\begin{split}
			\underbrace{\begin{bmatrix}
					M_1 & 0 \\
					0 & M_2
			\end{bmatrix}}_{\mathcal{M}^{(i)}}
			\underbrace{
				\begin{bmatrix}
					\partial_t\hat{x}_1 \\ \partial_t\hat{x}_2
			\end{bmatrix}}_{\dot{\mathcal{X}}^{(i)}} =
			\underbrace{\left(
				\begin{bmatrix}
					0 & P \\
					-P^T & 0
				\end{bmatrix} - 
				\begin{bmatrix}
					R_1 & 0 \\
					0 & R_2
				\end{bmatrix}
				\right)}_{\mathcal{J}^{(i)}-\mathcal{R}^{(i)}}
			\underbrace{\begin{bmatrix}
					\hat{e}_1 \\ \hat{e}_2
			\end{bmatrix}}_{\mathcal{E}^{(i)}} \\ +
			\underbrace{\begin{bmatrix}
					0 & B_1 \\
					B_2 & 0
			\end{bmatrix}}_{\mathcal{B}^{(i)}}
			\underbrace{\begin{bmatrix}
					(1-\beta)e_1^{(i-1)}(z_i) + \xi e_2^{(i-1)}(z_i)\\ -\beta e_1^{(i+1)}(z_{i+1}) + \xi e_2^{(i+1)}(z_{i+1}) \\ \beta e_2^{(i-1)}(z_i) + \tau e_1^{(i-1)}(z_i) \\ (\beta-1)e_2^{(i+1)}(z_{i+1}) + \tau e_1^{(i+1)}(z_{i+1})
			\end{bmatrix}}_{\mathcal{U}^{(i)}}
		\end{split}
	\end{align}
	with the (pointwise) discrete constitutive equation\footnote{$I_k$ represents the identity matrix of dimension $k$.}
	\begin{equation}
		\label{eq:dconst}
		\underbrace{\begin{bmatrix}
				\hat{e}_1 \\ \hat{e}_2
		\end{bmatrix}}_{\mathcal{E}^{(i)}} = 
		\underbrace{\begin{bmatrix}
				c_1I_{k_1} & 0 \\ 0 & c_2I_{k_2}
		\end{bmatrix}}_{\mathcal{Q}^{(i)}=(\mathcal{Q}^{(i)})^T>0}
		\underbrace{\begin{bmatrix}
				\hat{x}_1 \\ \hat{x}_2
		\end{bmatrix}}_{\mathcal{X}^{(i)}},
	\end{equation}
	the output $\mathcal{Y}^{(i)}=(\mathcal{B}^{(i)})^T\mathcal{E}^{(i)}$ and ${\mathcal{M}^{(i)}=(\mathcal{M}^{(i)})^T>0}$, ${\mathcal{J}^{(i)}=-(\mathcal{J}^{(i)})^T}$, ${\mathcal{R}^{(i)}=(\mathcal{R}^{(i)})^T\geq0}$
	is obtained.
\end{thm}

\begin{pf}
	Inserting our above numerical flux formulation and the approximation depending on the basis functions in \eqref{eq:strong} allows us to take all variables out of the integrals. Since ${\hat w_1\in\mathbb{R}^{k_1}}$ and ${\hat w_2\in\mathbb{R}^{k_2}}$ are arbitrary, we get the state space equations containing the matrices
	\begin{align}
		M_1 &= \int_{\Omega_i}\phi \phi^T\;\text{d}z \\
		M_2 &= \int_{\Omega_i}\psi \psi^T\;\text{d}z \\
		\begin{split}
			P &= \int_{\Omega_i}\partial_z\phi \psi^T\;\text{d}z \\ &+  \phi(z_i)(1-\beta)\psi^T(z_i) - \phi(z_{i+1})\beta\psi^T(z_{i+1}) 
		\end{split} \\
		R_1 &= \phi(z_i)\tau\phi^T(z_i) + \phi(z_{i+1})\tau\phi^T(z_{i+1}) \\
		R_2 &= \psi(z_i)\xi\psi^T(z_i) + \psi(z_{i+1})\xi\psi^T(z_{i+1}) \\
		B_1 &= \begin{bmatrix} \phi(z_i) & \phi(z_{i+1}) \end{bmatrix} \\
		B_2 &= \begin{bmatrix} \psi(z_i) & \psi(z_{i+1}) \end{bmatrix}.
	\end{align}
	The total energy of each element is given by
	\begin{equation}
		\label{eq:H_elem}
		H^{(i)} = \frac{1}{2}\int_{\Omega_i}c_1x_1^2 + c_2x_2^2\;\text{d}z.
	\end{equation}
	Inserting the approximations of state variables in \eqref{eq:H_elem} leads to the approximated Hamiltonian of each element,
	\begin{equation}
		\label{eq:H_approx_elem}
		\hat{H}^{(i)} = \frac{1}{2}\mathcal{X}^{(i)T}\mathcal{Q}^{(i)}\mathcal{M}^{(i)}\mathcal{X}^{(i)}.
	\end{equation}
	Since $c_1$ and $c_2$ are constants the discrete constitutive equation \citep{CardosoRibeiro2019}
	\begin{equation}
		\label{eq:gen_constitutive}
		\mathcal{M}^{(i)}\mathcal{E}^{(i)} = \partial_{\mathcal{X}^{(i)}}\hat{H}^{(i)}
	\end{equation}
	corresponds to \eqref{eq:dconst}.
	Taking the time derivative of \eqref{eq:H_approx_elem} gives us the output,
		\begin{equation}
			\frac{\text{d}\hat{H}^{(i)}}{\text{d}t} = - (\mathcal{E}^{(i)})^T\mathcal{R}^{(i)}\mathcal{E}^{(i)} + \underbrace{\mathcal{E}^{(i)T}\mathcal{B}^{(i)}}_{(\mathcal{Y}^{(i)})^T}\mathcal{U}^{(i)},
		\end{equation}
	which concludes the proof.
\end{pf}

\begin{rem}
	From \eqref{eq:gen_constitutive} it can be seen that the constitutive equation is evaluated for each element. Thus, this evaluation is much more cost efficient than the continuous Galerkin method \citep{Cardoso-Ribeiro2020}. This advantage might also be reflected for nonlinear Hamiltonian functions.
\end{rem}

\begin{assum}
	In the following, we will only consider nodal basis functions. These are formed from Lagrange polynomials. For example the discrete solution of a variable ${q(z,t)\in\mathbb{R}}$ in one element is given by
	\begin{equation}
		q^{(i)}(z,t) \approx \ell^T(z)\hat{q} = \sum_{j=0}^{k}\ell_j(z)\hat{q}_j(t),
	\end{equation}
	where ${\ell(z)}$ is the Lagrange polynomial with polynomial degree $k$ associated to element $i$ and node $j$ inside the element.
	Since Lagrange polynomials have the value one at node $j$ and zero at the other nodes, they simplify the interconnection of elements considerably compared to modal basis functions \citep{Hesthaven2008}.
\end{assum}

\subsection{Boundary conditions}
\label{sec:3.4}
So far, we focused on the interconnection of elements but did not consider the handling of boundary conditions. This raises the question of how to define the numerical fluxes at the boundaries of the integration domain $\Omega$. In the following, we introduce a particularly simple option for this. Since the boundary of the entire domain is split, ${\partial\Omega}=\Gamma_D\cup\Gamma_N$, the choice of the numerical flux depends on the boundary condition. Let us now consider ${z_{1}=a}$ (as $\Gamma_D$) for the left and ${z_{N+1}=b}$ (as $\Gamma_N$) for the right boundary. We specify the numerical fluxes for the desired boundary condition \eqref{eq:bc} as
	\begin{align}
		\label{eq:flux_DBC}
		e_1^{(1)*}(a) = u_1, \quad e_2^{(1)*}(a) = e_2^{(1)}(a)
	\end{align}
and
	\begin{align}
		\label{eq:flux_NBC}
		e_1^{(N)*}(b) = e_1^{(N)}(b), \quad e_2^{(N)*}(b) = u_2.
	\end{align}

\begin{rem}
	Applying the parameter set ${\beta=\tau=\xi=0}$, ${e_1^{(0)}=u_1}$ and ${e_2^{(N+1)}=u_2}$ for \eqref{eq:flux_a}-\eqref{eq:flux_d}
	at the boundaries leads to the above flux formulations \eqref{eq:flux_DBC} and \eqref{eq:flux_NBC}.
	Other ways of introducing boundary conditions can be found in the Literature and will be investigated in our further research.
\end{rem}


\subsection{Global system}
In order to generate the global system, we take the example boundary conditions of Section \ref{sec:3.4},
${e_1(a,t)=u_1}$ and ${e_2(b,t)=u_2}$, and demonstrate the global weak/"strong" DG form with inserted flux formulation
\begin{align}
	\small
	\begin{split}
		\sum_{i=1}^{N}\left(w_1, \partial_t x_1\right)_{\Omega_i} = \sum_{i=1}^{N}{\left(\partial_zw_1, e_2\right)_{\Omega_i}}
		\\ + \sum_{i=1}^{N-1}\left(\underbrace{ (\beta-1)w_1^{(i)}(z_{i+1})e_2^{(i+1)}(z_{i+1})}_{\text{\textcircled{1}}} + \underbrace{w_1^{(i)}(z_{i+1})\tau e_1^{(i+1)}(z_{i+1})}_{\text{\textcircled{2}}} \right)
		\\ + \sum_{i=1}^{N-1}\left({-\beta w_1^{(i)}(z_{i+1})e_2^{(i)}(z_{i+1})}-w_1^{(i)}(z_{i+1})\tau e_1^{(i)}(z_{i+1}) \right)
		\\ + \sum_{i=2}^{N}\left(\underbrace{\beta w_1^{(i)}(z_{i})e_2^{(i-1)}(z_{i})}_{\text{\textcircled{1}}} + \underbrace{w_1^{(i)}(z_{i})\tau e_1^{(i-1)}(z_{i})}_{\text{\textcircled{2}}} \right)
		\\ + \sum_{i=2}^{N}\left({(1-\beta)w_1^{(i)}(z_{i})e_2^{(i)}(z_{i})}-w_1^{(i)}(z_{i})\tau e_1^{(i)}(z_{i}) \right)
		\\ \underbrace{+ w_1^{(1)}(a)e_2^{(1)}(a)}_{\text{\textcircled{1}}} \underbrace{- w_1^{(N)}(b){u}_2}_{\text{\textcircled{3}}}
	\end{split} \\[2ex]
	\small
	\begin{split}
		\sum_{i=1}^{N}\left(w_2, \partial_t x_2\right)_{\Omega_i} = -\sum_{i=1}^{N}{\left(w_2, \partial_ze_1\right)_{\Omega_i}}
		\\ + \sum_{i=1}^{N-1}\left( \underbrace{-\beta w_2^{(i)}(z_{i+1})e_1^{(i+1)}(z_{i+1})}_{\text{\textcircled{1}}} + \underbrace{w_2^{(i)}(z_{i+1})\xi e_2^{(i+1)}(z_{i+1})}_{\text{\textcircled{2}}}\right)
		\\ + \sum_{i=1}^{N-1}\left({\beta w_2^{(i)}(z_{i+1})e_1^{(i)}(z_{i+1})}-w_2^{(i)}(z_{i+1})\xi e_2^{(i)}(z_{i+1}) \right)
		\\ + \sum_{i=2}^{N}\left(\underbrace{(1-\beta)w_2^{(i)}(z_{i})e_1^{(i-1)}(z_{i})}_{\text{\textcircled{1}}} + \underbrace{w_2^{(i)}(z_{i})\xi e_2^{(i-1)}(z_{i})}_{\text{\textcircled{2}}} \right)
		\\ + \sum_{i=2}^{N}\left({ (\beta-1)w_2^{(i)}(z_{i})e_1^{(i)}(z_{i})}-w_2^{(i)}(z_{i})\xi e_2^{(i)}(z_{i}) \right)
		\\ \underbrace{- w_2^{(1)}(a)e_1^{(1)}(a)}_{\text{\textcircled{1}}} \underbrace{+ w_2^{(1)}(a){u}_1}_{\text{\textcircled{3}}}.
	\end{split}
\end{align}
Following Theorem \ref{thm:fem} for each element gives us the global finite-dimensional PH state space model
	\begin{equation}
		\label{eq:final_phs}
		\mathcal{M\dot{X}}=(\mathcal{J-R})\mathcal{E}+\mathcal{GU}
	\end{equation}
where
	\begin{align}
		\mathcal{X}^T &= \begin{bmatrix}
			(\mathcal{X}^{(1)})^T & \dots &(\mathcal{X}^{(N)})^T
		\end{bmatrix}
		\\
		\mathcal{E}^T &= \begin{bmatrix}
			(\mathcal{E}^{(1)})^T & \dots &(\mathcal{E}^{(N)})^T
		\end{bmatrix}
		\\
		\mathcal{M} &= \mathrm{diag}(\mathcal{M}^{(1)},\dots,\mathcal{M}^{(N)})=\mathcal{M}^T>0 \\
		\mathcal{J} &= \mathrm{diag}(\mathcal{J}^{(1)},\dots,\mathcal{J}^{(N)}) + \mathcal{F}=-\mathcal{J}^T \\
		\mathcal{R} &= \mathrm{diag}(\mathcal{R}^{(1)},\dots,\mathcal{R}^{(N)})-\mathcal{D}=\mathcal{R}^T\geq0
	\end{align}
The skew-symmetric matrix $\mathcal{F}$ results form the numerical flux interconnecting the elements and fulfilling the boundary conditions, see \textcircled{1}. ${\mathcal{D}=\mathcal{D}^T}$ is part of the numerical damping matrix $\mathcal{R}$, see \textcircled{2}. The vector $\mathcal{GU}$ results from \textcircled{3}. In order to close the system representation, the constitutive equation \eqref{eq:dconst} of each element is required, which leads to
	\begin{equation}
		\mathcal{E} = \underbrace{\mathrm{diag}(\mathcal{Q}^{(1)},\dots,\mathcal{Q}^{(N)})}_{\mathcal{Q}=\mathcal{Q}^T>0}\mathcal{X}.
	\end{equation}
The total energy is given by
	\begin{equation}
		\label{eq:H_approx_total}
		\hat{H} = \sum_{i=1}^N\hat{H}^{(i)}=\frac{1}{2}\mathcal{X}^T\mathcal{Q}\mathcal{M}\mathcal{X}.
	\end{equation}
and its time derivative
	\begin{equation}
		\frac{\text{d}\hat{H}}{\text{d}t} = \underbrace{\mathcal{E}\mathcal{G}}_{\mathcal{Y}^T}\mathcal{U} - \mathcal{E}^T\mathcal{RE}
	\end{equation}
representing the power transmitted through the boundary ${\partial\Omega}$ plus dissipation due to numerical damping gives us the power conjugated output $\mathcal{Y}$.

With a state transformation equivalently to \cite{Cardoso-Ribeiro2020},
	\begin{equation}
		\label{eq:transformation}
		\tilde{\mathcal{X}} = \mathcal{M}\mathcal{X},
	\end{equation}
inserted in \eqref{eq:H_approx_total},
	\begin{equation}
		\hat{H} = \frac{1}{2}\tilde{\mathcal{X}}^T\mathcal{M}^{-T}\mathcal{Q}\tilde{\mathcal{X}},
	\end{equation}
we get
	\begin{equation}
		\mathcal{E} = \partial_{\tilde{\mathcal{X}}}\hat{H},
	\end{equation}
and obtain the "classical" input-state-output PH state space model
	\begin{equation}
		\dot{\tilde{\mathcal{X}}} = \mathcal{J}\partial_{\tilde{\mathcal{X}}}\hat{H} + \mathcal{G}\mathcal{U},
	\end{equation}
which contains the original output
	\begin{equation}
		\mathcal{Y} = \mathcal{G}^T\partial_{\tilde{\mathcal{X}}}\hat{H}.
	\end{equation}

\begin{rem}
	Due to the block diagonal form of $\mathcal{M}$ the state transformation can be computed very efficiently.
\end{rem}

\section{Numerical results}
\label{sec:4}
In this section, the performance of our approach is demonstrated by a simple FEniCS \citep{LoggMardalEtAl2012} simulation. Therefore, we apply the structure preserving DG method on an example system (hyperbolic wave equation).

The wave equation
	\begin{align}
		-\begin{bmatrix}
			\dot{p} \\ \dot{q}
		\end{bmatrix} =
		\begin{bmatrix}
			0 & \partial_z \\
			\partial_z & 0
		\end{bmatrix}
		\begin{bmatrix}
			e_p \\ e_q
		\end{bmatrix}
	\end{align}
with the states ${p(z,t)\in\mathbb{R}}$ and ${q(z,t)\in\mathbb{R}}$ is defined on the domain $\Omega=[0,1]$. With the Hamiltonian
	\begin{equation}
		H = \frac{1}{2}\int_{0}^{1} p^2 + q^2 \;\text{d}z
	\end{equation}
the efforts are ${e_p(z,t)=p(z,t)}$ and ${e_q(z,t)=q(z,t)}$. In order to form the boundary value problem, we consider the initial
	\begin{align}
		p(z,0) &= 0 \\
		q(z,0) & = 0
	\end{align}
and the boundary conditions
	\begin{align}
		u_1(t) = e_p(0,t) &= \begin{cases}
			\sin(8\pi t), & \quad 0\leq t < 0.125\\
			0, & \quad t \geq 0.125
		\end{cases} \\
		u_2(t) = e_q(1,t) & = 0.
	\end{align}
This leads to a traveling wave of $p$ and $q$ from left to right and back.

\subsection{Simulations}

The wave equation is discretized with ${N=50}$ equidistant elements using first order Lagrange polynomials for $\phi$ and $\psi$ leading to ${\hat{p}\in\mathbb{R}^{100}}$ and ${\hat{q}\in\mathbb{R}^{100}}$. The benchmark system is simulated for ${T=1.5}$ where the implicit midpoint rule with the sampling time ${\Delta T = 2.5\cdot 10^{-4}}$ is used for solving the ordinary differential equation \eqref{eq:final_phs}. We run three simulations containing different numerical flux parameter sets listed in Table \ref{tab:sim}.

\begin{table}[htbp]
	\centering
	\caption{Numerical fluxes for simulation}
	\label{tab:sim}
	\begin{tabular}{ l | l }
		Numerical flux & Parameters  \\
		\hline
		Central flux & $\beta=0.5$, $\tau=\xi=0$ \\
		Upwind flux & $\beta=0$, $\tau=\xi=0$ \\
		Damped central flux &  $\beta=0.5$, $\tau=\xi=0.5$ \\
		\hline
	\end{tabular}
\end{table}

\begin{figure}[htbp]
	\centering
	\input{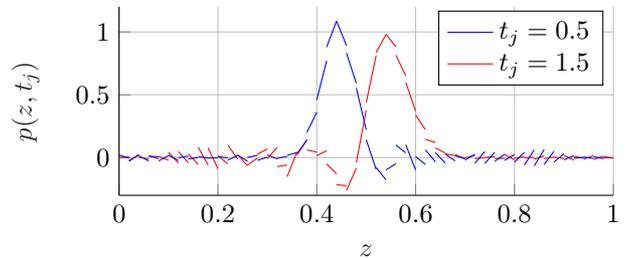}
	\caption{Numerical solution of $p$ due to central flux}
	\label{fig:central_p}
\end{figure}

\begin{figure}[htbp]
	\centering
	\input{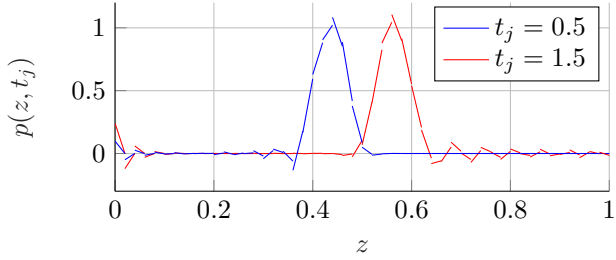}
	\caption{Numerical solution of $p$ due to upwind flux}
	\label{fig:upwind_0_p}
\end{figure}

\begin{figure}[htbp]
	\centering
	\input{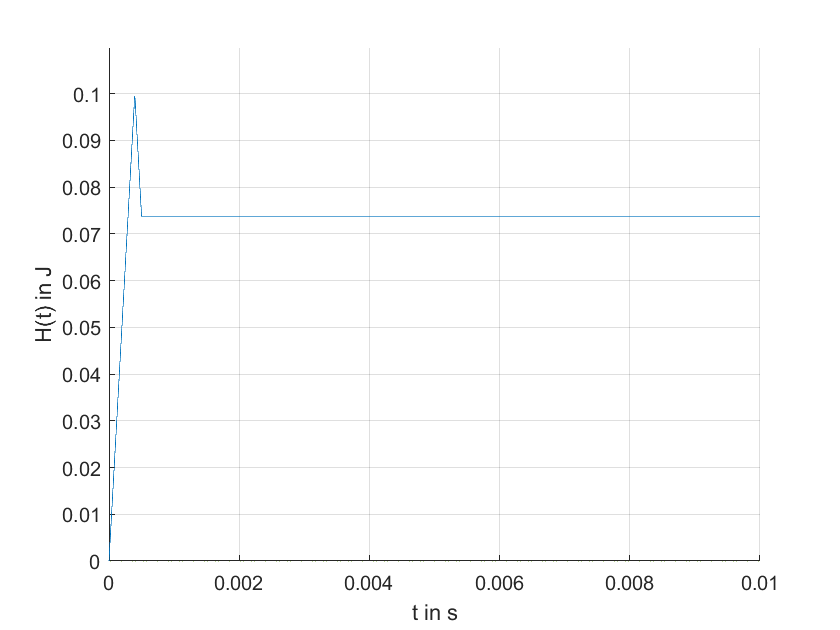}
	\caption{Hamiltonian $H$ due to different numerical fluxes}
	\label{fig:H}
\end{figure}

Fig. \ref{fig:central_p}-\ref{fig:damped_central_p} represent numerical solutions of $p$ at two time steps ${t_j=\{0.5,1.5 \}}$ (one each before and after the reflection) for different fluxes. Compared to the upwind flux, the central flux performs worse and leads to large discontinuities between elements. Using the damped central flux the discontinuities are further reduced by means of numerical damping. The effect of numerical damping also appears in the Hamiltonian. Fig. \ref{fig:H} shows the total energy of the system which keeps a constant level at ${t\geq0.125}$ for flux formulations where ${\tau=\xi=0}$. In case of the damped central flux energy is taken from the system.

\begin{figure}[htbp]
	\centering
	\input{./fig/damped_central_p}
	\caption{Numerical solution of $p$ due to damped central flux}
	\label{fig:damped_central_p}
\end{figure}

\begin{rem}
	To illustrate the discontinuity between the elements in the figures, we have used a small number of elements.
	The accuracy of the solution naturally increases with an increasing number of elements. In addition, the effect of dissipation due to numerical damping is reduced, since this is caused by the discontinuous behavior between the elements.
	E.g., when the number of elements is quadrupled (${N=200}$), there is hardly any difference in the Hamiltonian of Figure \ref{fig:H} compared to the conservative systems, where ${\tau=\xi=0}$.
\end{rem}

\subsection{Eigenvalue spectrum}
\begin{figure}[htbp]
	\centering
	\input{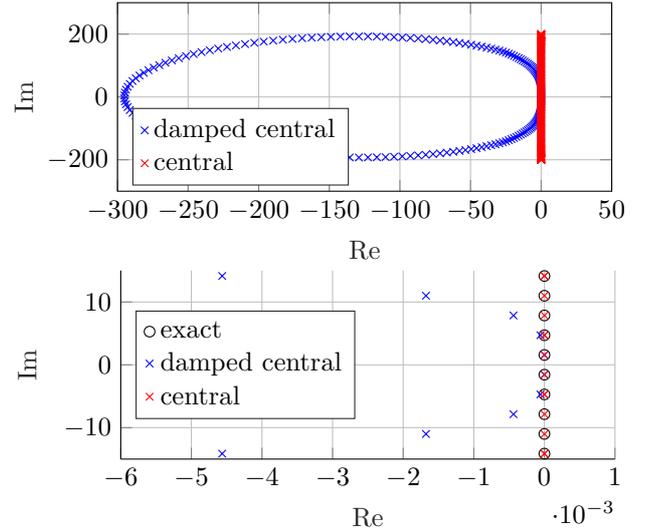}
%
%
\definecolor{mycolor1}{rgb}{0.00000,0.44700,0.74100}%
\definecolor{mycolor2}{rgb}{0.85000,0.32500,0.09800}%
\begin{tikzpicture}

\begin{axis}[%
width=6.5cm,
height=2.5cm,
scale only axis,
xmin=-0.006,
xmax=0.001,
xlabel style={font=\color{white!15!black}},
xlabel={Re},
ymin=-15,
ymax=15,
ylabel style={font=\color{white!15!black}},
ylabel={Im},
axis background/.style={fill=white},
axis x line*=bottom,
axis y line*=left,
xmajorgrids,
ymajorgrids,
legend style={at={(0.03,0.5)}, anchor=west, legend cell align=left, align=left, draw=white!15!black}
]
\addplot [color=black, only marks, mark=o, mark options={solid, black}]
  table[row sep=crcr]{%
0	1.5707963267949\\
0	4.71238898038469\\
0	7.85398163397448\\
0	10.9955742875643\\
0	14.1371669411541\\
0	17.2787595947439\\
0	20.4203522483337\\
0	23.5619449019234\\
0	26.7035375555132\\
-0	-1.5707963267949\\
-0	-4.71238898038469\\
-0	-7.85398163397448\\
-0	-10.9955742875643\\
-0	-14.1371669411541\\
-0	-17.2787595947439\\
-0	-20.4203522483337\\
-0	-23.5619449019234\\
-0	-26.7035375555132\\
};
\addlegendentry{exact}

\addplot [color=blue, only marks, mark=x, mark options={solid, blue}]
  table[row sep=crcr]{%
-0.0874135330161177	29.8604040021529\\
-0.0874135330161177	-29.8604040021529\\
-0.129210002530924	33.011403515739\\
-0.129210002530924	-33.011403515739\\
-0.0345406017495244	23.5667997292736\\
-0.0345406017495244	-23.5667997292736\\
-0.0565224364373773	26.7124621681786\\
-0.0565224364373773	-26.7124621681786\\
-0.00167743400777765	10.9956870236234\\
-0.00167743400777765	-10.9956870236234\\
-0.0101277683678984	17.2798186022361\\
-0.0101277683678984	-17.2798186022361\\
-0.0196300633021962	20.4227620209588\\
-0.0196300633021962	-20.4227620209588\\
-0.00456341382473102	14.1375595211051\\
-0.00456341382473102	-14.1375595211051\\
-0.000438129425225808	7.85400273653069\\
-0.000438129425225808	-7.85400273653069\\
-5.69109737571516e-05	4.71239062870776\\
-5.69109737571516e-05	-4.71239062870776\\
-7.03409252511555e-07	1.57079633359343\\
-7.03409252511555e-07	-1.57079633359343\\
};
\addlegendentry{damped central}

\addplot [color=red, only marks, mark=x, mark options={solid, red}]
table[row sep=crcr]{%
	2.66453525910038e-15	30.0648580463559\\
	2.66453525910038e-15	-30.0648580463559\\
	5.55111512312578e-15	26.8611923712035\\
	5.55111512312578e-15	-26.8611923712035\\
	-7.99360577730113e-15	32.8209649195443\\
	-7.99360577730113e-15	-32.8209649195443\\
	2.3037127760972e-15	20.4910469788004\\
	2.3037127760972e-15	-20.4910469788004\\
	-9.54791801177635e-15	17.3216344906173\\
	-9.54791801177635e-15	-17.3216344906173\\
	5.62050406216485e-15	1.57082862430239\\
	5.62050406216485e-15	-1.57082862430239\\
	3.52495810318487e-15	7.85801672363831\\
	3.52495810318487e-15	-7.85801672363831\\
	6.43929354282591e-15	11.0066407798899\\
	6.43929354282591e-15	-11.0066407798899\\
	2.22044604925031e-16	23.5014632276614\\
	2.22044604925031e-16	-23.5014632276614\\
	-2.66453525910038e-15	33.2828293297358\\
	-2.66453525910038e-15	-33.2828293297358\\
	-5.32907051820075e-15	23.6704067771399\\
	-5.32907051820075e-15	-23.6704067771399\\
	-2.91433543964104e-15	14.1240921941296\\
	-2.91433543964104e-15	-14.1240921941296\\
	-1.99840144432528e-15	14.1606707371214\\
	-1.99840144432528e-15	-14.1606707371214\\
	-5.55111512312578e-16	4.71190453211518\\
	-5.55111512312578e-16	-4.71190453211518\\
	-1.28785870856518e-14	4.71326086227819\\
	-1.28785870856518e-14	-4.71326086227819\\
};
\addlegendentry{central}

\end{axis}
\end{tikzpicture}%
	\caption{Eigenvalue spectrum}
	\label{fig:Eigenvalue}
\end{figure}

In this sections, we apply the same discretization parameters as above to calculate the eigenvalues $\lambda(\mathcal{M}^{-1}\mathcal{J})$ of the discrete PH systems. With the conservative numerical fluxes, we obtain only complex eigenvalues (similar to the exact ones), whereas the numerical damping shifts the eigenvalues to the right complex half plane, see Figure \ref{fig:Eigenvalue}.

\section{Conclusion}
\label{sec:5}
We presented a DG discretization procedure for one-dimensional PH systems of conservation laws in order to achieve a PH state space model. The resulting finite-dimensional system preserves the PH structure containing a block-diagonal mass matrix which allows to calculate the constitutive equation efficiently. Since the mixed boundary conditions are applied in a weak sense, classic control or model order reduction schemes can be easily applied to the explicit PH state space model.

Since we have so far only presented a first concept for the structure preserving discretization of PH systems using DG methods (introducing numerical damping, whose influence decreases with a large number of elements), we would like to elaborate this further in the future.
We are currently working  on the extension for multidimensional nonlinear PH systems with the possibility to include more complex numerical fluxes.


\bibliography{mybib_new}             


\end{document}